\documentclass[11pt,a4paper]{article}
\usepackage{graphicx}
\usepackage{tabularx}
\usepackage{multicol}
\usepackage{tabto}
\usepackage{url}
\usepackage{enumitem}
\usepackage{textcomp}
\usepackage{eurosym}
\usepackage{sectsty}
\usepackage{xcolor}
\usepackage{booktabs}
\usepackage{colortbl}
\usepackage{fancyhdr}
\usepackage[yyyymmdd]{datetime}
\usepackage[font=sf]{caption}

\newcolumntype{L}[1]{>{\raggedright\arraybackslash}p{#1}}
\newcolumntype{C}[1]{>{\centering\arraybackslash}p{#1}}
\newcolumntype{R}[1]{>{\raggedleft\arraybackslash}p{#1}}
\newcolumntype{S}[1]{>{\raggedright\arraybackslash}p{#1}}

\usepackage{flowchart}
\usetikzlibrary{shapes,arrows.meta,chains,external,positioning}

\newcommand{\eat}[1]{}

\newcommand{\ts}{
    \tikzset{>={Latex[width=2mm,length=2mm]}}
    \tikzstyle{line} = [draw, ->, >=latex, ultra thick]
    \tikzstyle{box} = [
      draw,
      rectangle,
      thick,
      align=center,
      text width=2.5cm,
      text centered,
      minimum height=4.5em
    ]
    \tikzstyle{circ} = [
      circle,
      align=center,
      text width=4em,
      text centered,
      inner sep=1mm,
      outer sep=1mm,
      minimum width=0cm,
      minimum height=0cm
    ]
    \tikzstyle{noshape} = [text width=5em, text centered, minimum height=5em]
}

\begin{document}
\thispagestyle{empty}

\bibliographystyle{plain}

\begin{center}
\end{center}

\NumTabs{6}

\sf

\begin{center}

\begin{tabular}{|p{166mm}|}\hline
\begin{center}{\Huge

Emergency Financing Tokens\\

\vspace{5pt}}
{\large Geoffrey Goodell (goodell@dts-ltd.co.uk)}
\vspace{5pt}\end{center}

{\large

We propose a novel payment mechanism for use by victims of large-scale conflict
or natural disasters to conduct critical economic transactions and rebuild
damaged infrastructure in the absence of both cash and traditional electronic
payment mechanisms linked to bank accounts, such as debit cards or wire
transfers.  Claimants shall receive electronic tokens that can be used to pay
registered businesses, such as purveyors of food and other basic goods,
providers of essential services, and contractors to carry out construction
tasks.  The system shall be based upon the scalable architecture for retail
payments described in our earlier work~\cite{goodell2022}, which provides both
strong privacy for consumers and strong compliance enforcement for recipients
of funds.  The system shall be designed to achieve three main objectives.
First, tokens issued to claimants would be held directly by the claimants
themselves, not via intermediaries, to avoid the risk of failure or subversion
of asset custodians.  Second, transactions shall not be traceable to the
identity of the claimants, thus mitigating the risk that claimants can be
pressured by service providers or other parties to reveal information that can
be used to exploit them.  Third, businesses and service providers that receive
tokens shall be subject to rigorous compliance procedures upon redemption for
cash or bank deposits, thus ensuring that only legitimate businesses or service
providers can receive value from tokens, that token transfers will embed the
identities of any recipients beyond the initial claimant, and that tax
obligations shall be met at the time of redemption.

\vspace{10pt}}\\

\hline\end{tabular}
\end{center}

\sectionfont{\normalsize\noindent\sf\textbf}

\section*{SUMMARY}

For many people, the twenty-first century is a turbulent time, fraught with
risk.  Natural disasters resulting from climate change as well as human
conflicts such as the war in Ukraine cause substantial destruction and
displacement.  During large-scale displacement events, individuals may require
emergency financing to ensure that they have adequate funds to pay for
essential goods and services, such as food, water, medicine, and clothing.
They also require funds to rebuild destroyed homes.  But allocating funds is
not enough: victims of disasters need a means to pay for goods and services, in
a manner that is both private and secure.  Prevailing electronic payment
methods are linked to banks and payment networks, which themselves might be
inoperable, unreachable, or compromised by state or non-state actors.  Users
might be particularly vulnerable to surveillance or expropriation of their
personal data associated with card payments or bank transfers, and the stakes
can be higher than usual.  On the other hand, cash introduces a different set
of risks during a crisis.  Its distribution, which generally requires bank
branches, ATMs, specialised vehicles, and transportation networks, can easily
be interrupted or attacked.  Even when cash is available, it can be used
without regulatory oversight, inviting the risk of use by illegitimate actors
for unscrupulous activities without the usual human-level checks that might be
available during normal times.

As part of disaster relief and recovery efforts, emergency funding might be
sourced via (1) government grants, (2) private-sector donations, or (3)
private-sector investments.  In the third case, funding might be backed by (1)
a bond or other debt security issued by an institution or special investment
vehicle; (2) illiquid assets transferred by the government, such as seized
assets; or (3) the personal illiquid assets of the claimant, such as the
claimant's home or other real property.

Our proposed system allows consumers to file claims in exchange for token
assets that they can spend without associating their identities with their
payments, thus protecting the consumers from blackmail or unwanted profiling.
The tokens are self-validating and make use of a system of relays to ensure
their integrity and uniqueness.  When a consumer makes a payment, the consumer
embeds the identity of the recipient into the token, thus ensuring that when
the token is eventually redeemed, compliance rules specifying obligations for
the recipient can be enforced.  Such compliance rules can include tax
enforcement, AML/KYC checks, anti-fraud measures, and restrictions that
constrain how the recipient can transact the token onward.  The combination of
strong consumer privacy with strong compliance enforcement for vendors makes
the system particularly suitable for emergency scenarios, wherein consumers are
at risk of profiling, surveillance, and discrimination, while corrupt or
unscrupulous businesses might seek to evade rules that dictate the terms under
which they are allowed to operate.

\section*{HOW IT WORKS}

\begin{figure}[ht]
\begin{center}
\hspace{-0.8em}\scalebox{1}{
\begin{tikzpicture}[>=latex, node distance=3cm, font={\sf \small}, auto]\ts
\tikzset{>={Latex[width=4mm,length=4mm]}}
\node (w1) at (-0.8, 0) [] {
    \scalebox{0.08}{\includegraphics{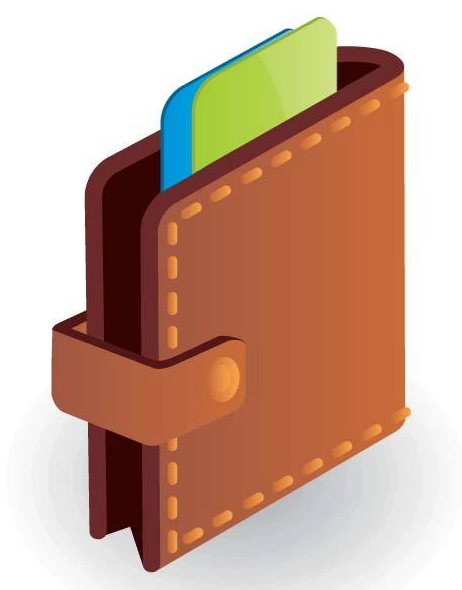}}
};
\node (b1) at (4, 6) [] {
    \scalebox{2}{\includegraphics{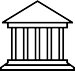}}
};
\node (s1) at (4, -6) [] {
    \scalebox{2}{\includegraphics{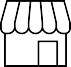}}
};
\node (r1) at (3.4, -1.8) [] {
    \scalebox{0.3}{\includegraphics{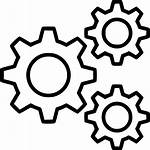}}
};
\node (r2) at (3.4, 1.8) [] {
    \scalebox{0.3}{\includegraphics{images/gears.jpg}}
};
\node (r3) at (7.0, -1.8) [] {
    \scalebox{0.3}{\includegraphics{images/gears.jpg}}
};
\node (r4) at (7.0, 1.8) [] {
    \scalebox{0.3}{\includegraphics{images/gears.jpg}}
};
\node (c0) at (5.3,0) [
    circ,
    draw,
    line width=1.2mm,
    color=orange,
    minimum height=3.5cm,
    minimum width=3.5cm
] {};
\node (c1) at (5.3,0) [circ, draw, thick, minimum height=3.4cm, minimum width=3.4cm] {};
\node (c2) at (5.3,0) [circ, draw, thick, minimum height=3.6cm, minimum width=3.6cm] {};
\node (p1) at (5.3,0) [align=center] {\textbf{Integrity}\\DLT System\\(relays)};
\node (x1) at (10.4, 3.4) [] {
    \scalebox{2}{\includegraphics{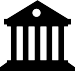}}
};
\node (b3) at (10.4, -3.4) [] {
    \scalebox{2}{\includegraphics{images/s-sender-bank.png}}
};
\node (i1) at (-2.0, 0) [noshape, text width=2cm, align=center] {consumer wallet};
\node (i2) at (2.4, 6) [noshape, text width=1.8cm, align=center] {distribution agent};
\node (i4) at (2.6, -6) [noshape, text width=1.8cm, align=center] {merchant};
\node (i5) at (12.0, -3.4) [noshape, text width=1.8cm, align=center] {redemption agent};
\node (i6) at (11.8, 3.4) [noshape, text width=1.8cm, align=center] {issuer};
\draw[->, line width=1mm] (b1) edge[bend right=20] node[above, sloped] {receive} (w1);
\draw[->, line width=1mm] (x1) edge[bend right=20] node[above, sloped] {distribute} (b1);
\draw[->, line width=1mm] (b3) edge[bend right=20] node[above, sloped] {redeem} (x1);
\draw[->, line width=1mm] (w1) edge[bend right=20] node[above, sloped] {spend} (s1);
\draw[<->, line width=1mm] (s1) edge[bend right=20, color=red] (r1);
\draw[<->, line width=1mm] (b1) edge[bend left=20, color=red] (r2);
\draw[<->, line width=1mm] (b3) edge[bend right=20, color=red] (r3);
\draw[->, line width=1mm] (s1) edge[bend right=20] node[above, sloped] {return} (b3);
\draw[<->, line width=1mm] (x1) edge[bend left=20, color=red] (r4);
\end{tikzpicture}}
\end{center}

\caption{Schematic representation of a system for issuing and redeeming digital tokens.}
\label{f:overview}

\end{figure}

Once a claim for emergency finance is approved, claimants shall use a
\textit{bearer wallet} (``non-custodial wallet'') to generate a set of tokens
with suitable denominations determined by the issuer.  Please refer to
Figure~\ref{f:overview}.  The tokens shall embed the identity of a
\textit{relay provider} that can optionally be a local notary or bank, or the
issuer itself.  The claimant shall use a blind signature
scheme~\cite{chaum1982,chaum2021} to request accreditation from the issuer for
each of the tokens.  The resulting unblinded tokens can be used to make
payments that protect the privacy of the claimant.  Then, the claimant has the
task of finding an authorised vendor who can perform qualifying services.
Vendors can identify themselves to claimants with certificates signed by the
issuer, which claimants shall be able to verify using the public key of the
issuer.  The claimant shall then sign tokens over to a vendor using the
identity key of that vendor, registering the update with the relay provider
either directly or via a point of sale device or service provided by the
vendor~\cite{goodell2022}.  The relay provider shall then provide a proof of
provenance demonstrating that the record either has been added or shall be
added to the official history trusted by the issuer~\cite{coward2022}.  The
official history can be maintained directly by the issuer itself or via a
(permissioned) distributed ledger operated by relays overseen by funding
organisations~\cite{goodell2022a}.  The proof of provenance thus allows the
vendor to accept the payment from the claimant.  Optionally, the system design
can either allow or disallow the vendor to subsequently transfer tokens to
other authorised vendors, who might, for example, serve as suppliers or
subcontractors.  Finally, authorised vendors shall be able to transform spent
tokens to bank deposits or cash by transferring the tokens to a local bank,
which can verify the validity and integrity of the tokens as a condition of
acceptance, and redeem the tokens with the issuer in exchange for the funds.

Our proposed system does not require specially manufactured hardware or
certified devices to function.  We anticipate that the tools used by consumers,
merchants, and token issuers will be run on commodity laptops (\texttt{amd64}
architecture or equivalent) or mobile devices (phones or tablets with
\texttt{aarch64} architecture or equivalent), running a Linux-based operating
system and using a low-cost, non-broadband Internet connection.  For the
purposes of pilots that can be launched in the first year, the storage,
network, and computation requirements for infrastructure devices (specifically
relays) are compatible with the use of consumer-grade equipment.  To protect
their assets, users may, at their option, choose devices that offer security
features such as encryption, passcodes, or local biometrics to protect against
theft, and they may make copies of their tokens to mitigate the risk of
accidental loss of their devices.

\section*{PROJECT TIMETABLE AND DELIVERABLES}

We have requested USD \$1.4M to be used to support the first twelve months of
delivery of services by a London-based business entity to The Peace Coalition
in support of its plan for rebuilding Ukraine.  Our services shall mainly
involve the delivery of a live implementation of Emergency Financing Tokens and
the operation of a team and infrastructure necessary to support it.  The first
twelve months shall mostly involve the development and evaluation of a working
technical process along with open-source software to support that process,
based upon software published by University College London under the
three-clause BSD licence~\cite{comet}.

\noindent Our work items shall include:

\begin{enumerate}

\item\label{i:issuer} A straightforward \textit{issuer} application to
facilitate creating and redeeming blinded tokens, in a manner similar to what
has been proposed by BIS Project Tourbillon~\cite{bis2022}.  This system
shall employ the blind signature mechanisms in COMET, and we shall leverage
security professionals to rigorously evaluate the security of this system.

\item\label{i:consumer} A straightforward \textit{consumer} application
designed to interact with the simple issuer application, combined with a
mechanism for paying merchants in real money, externally to our system.

\item\label{i:merchant} A \textit{merchant} application facilitating blinded
tokens to be locked and validated by the issuing party, so that they can
effectively be spent.  Presumably, this would entail having the vendor provide
a channel by which the consumer would be able to contact the issuer and furnish
the token along with a request to transfer the token to the vendor.  The
request would be signed by the holder of the token; presumably this would be
possible because the token itself would be a public key corresponding to a
private key held exclusively by the consumer.

\item\label{i:relay} A \textit{relay} application enabling the consumer to
transfer a token to the merchant without involving the issuer, and to avoid
requiring the issuer to hold a record of all of the tokens that it had
validated.  This system shall exercise most of the

\item\label{i:ledger} A permissioned distributed ledger system that distributes
the role of the relay system across multiple, independent entities, providing a
means by which the issuer can trust the shared history constructed by consensus
of the relays, without having to trust the version of the truth provided by any
particular relay.  This mechanism allows the system that facilitates
transactions to achieve greater scale without being operated by a central
party.

\end{enumerate}

Note that our set of work items do not include a \textit{minting} system, which
would allow tokens to circulate without being reissued~\cite{goodell2022}.
This role is important for full-featured digital currency systems, wherein
money would circulate through the economy without returning to the issuer.
However, is not necessary to achieve the outcomes described in this proposal.

\section*{ACKNOWLEDGEMENTS}

We acknowledge University College London, the UCL Future of Money Initiative,
and Professor Tomaso Aste for their continued support of our work.

\section*{RELATED ARTICLES}

\begin{itemize}

\item G Goodell, H Al-Nakib, and P Tasca.  ``A Digital Currency Architecture
for Privacy and Owner-Custodianship.''  \textit{Future Internet} 2021, 13(5),
130, May 2021, \url{https://doi.org/10.3390/fi13050130}.

\item A Rychwalska, G Goodell, and M Roszczynska-Kurasinska.  ``Data management
for platform-mediated public services: Challenges and best practices.''
\textit{Surveillance \& Society}, Vol 19 No 1 (2021), pp. 22-36.
\url{https://doi.org/10.24908/ss.v19i1.13986}.

\item G Goodell, H Al-Nakib, and P Tasca.  ``Digital Currency and Economic
Crises: Helping States Respond.''  LSE Systemic Risk Centre Special Papers SP
20, September 2020, \url{https://ssrn.com/abstract=3622089}.

\item G Goodell and T Aste.  ``Can Cryptocurrencies Preserve Privacy and Comply
with Regulations?''  \textit{Frontiers in Blockchain}, May 2019.
\url{https://doi.org/10.3389/fbloc.2019.00004}.

\item K Coward and D Toliver.  ``Rigging Specifications.''  Technical
specification, August 2022. \url{http://trie.site/rigging_specifications.pdf}

\end{itemize}

\end{document}